\documentclass[12pt,psfig,epsfig]{article}
\newcommand{\beq}{\begin{equation}}
\newcommand{\eeq}{\end{equation}}
\newcommand{\bea}{\begin{eqnarray}}
\newcommand{\eea}{\end{eqnarray}}
\begin{document}
\begin{center}{\large {\bf Constraining Galactic $p\gamma$ Interactions with Cosmic Ray Electron and Positron Spectra}} 

\medskip
{Nayantara Gupta\footnote{Email: nayan@physics.unlv.edu}, 
Bing Zhang\footnote{Email: bzhang@physics.unlv.edu}}\\
 {Department of Physics, 
University of Nevada Las Vegas, Las Vegas, NV 89154, USA}
\end{center}

\begin{abstract}
High energy protons produced by various sources of cosmic rays, {\it e.g.} supernova remnants, pulsar wind nebulae, active galactic nuclei and gamma-ray bursts participate in $p\gamma$ and $pp$ interactions.  Although
$pp$ interactions may be the dominant mechanism in our Galaxy, it is
unclear how important $p\gamma$ process is. 
We show that the upper bound on the fraction of  
protons participating in $p\gamma$ interactions inside all Galactic
astrophysical sources of cosmic rays is $10\%$. 
\end{abstract}


\date{\today}

\section{Introduction:}
High energy protons are expected to be emitted by various astrophysical objects {\it e.g.} supernova remnants (SNRs), pulsar wind nebulae (PWN), microquasars,
magnetars, active galactic nuclei (AGN) and gamma-ray bursts
(GRBs). The relativistic protons interact with other protons and photons. High energy neutrinos and photons are subsequently produced in these interactions. Protons and other charged particles are deflected by Galactic and extra-galactic magnetic field before they arrive us. It is possible to trace the sources of cosmic rays by detecting neutral particles {\it e.g.} neutrinos, photons from them. High energy neutrinos cannot be produced inside astrophysical sources of cosmic rays if they contain only leptons. Hence, detection of high energy neutrinos from a source would also be a signature of the existence of baryons inside that source. 
A large number of neutrino telescopes are under operation/construction
to explore the neutrino universe in the near future. Innovative
detection methods have been employed in building these neutrino
detectors \cite{amanda1,baikal,antares,K3net,ice1,rice,anita,lofar}. 
AMANDA, the high energy muon and neutrino detector in operation near
the south pole has put limits on the neutrino fluxes from various
point sources
\cite{amanda2,amanda1} in the northern hemisphere. The current limit
on the diffuse all flavor neutrino flux is
$3.3\times10^{-7}{\rm GeVcm^{-2}sec^{-1}sr^{-1}}$ in the energy range of
16TeV to 2PeV \cite{amanda3}. This limit is from the data taken by
AMANDA-II between 2000 to 2003. In near future a more conservative
limit of $4.2\times10^{-9}{\rm GeVcm^{-2}sec^{-1}sr^{-1}}$ or a real
detection of the neutrino background is expected from
IceCube experiment after three years of operation \cite{ice2}.

Hadrons and leptons are expected to be shock accelerated inside
astrophysical objects by the Fermi mechanism to very high
energies. Charged and neutral pions can be produced in proton photon,
proton proton or nucleon nucleon interactions, and the charged pions
subsequently decay to produce neutrinos. Neutrino production
interactions are $p\gamma\rightarrow
\pi^{+}n\rightarrow e^{+}\nu_{\mu}\nu_{e}\bar\nu_{\mu}n$ and
$pp\rightarrow \pi^{\pm}\rightarrow
e^{\pm}\nu_{\mu}\nu_{e}(\bar\nu_e)\bar\nu_{\mu}$. We assume $pn$
interactions are similar to $pp$ interactions. The neutral pions decay
to $\gamma$-rays. It is difficult to know what fractions of all protons
 in our Galaxy participate in photo-pion ($p\gamma$) production and 
$pp$ interactions. The positrons lose energy during their propagation unlike neutrinos. The propagation of electrons and positrons in our Galaxy has been studied earlier \cite{strong}. It has been suggested \cite{anch} that the detection of Glashow resonance \cite{glash} $\bar\nu_e$ events at 6.3PeV by IceCube detector may offer us an opportunity to know whether $pp$ interactions are the
dominant mechanisms of ultrahigh energy neutrino production in the
universe. The ratios of $\bar\nu_e$ to muon flavor neutrinos
($\bar\nu_e:(\nu_{\mu}+\bar\nu_{\mu})$) at the Glashow resonance
energy (6.3 PeV) would be different for neutrinos produced in
$p\gamma$ and $pp$ interactions, and by measuring these ratios it would
also be possible to estimate neutrino mixing angles at ultrahigh
energies \cite{pijush}. However, effects like energy loss by muons is
expected to reduce \cite{kashti} and the decay of $\mu^{-}$ and
$\mu^{+}$ produced in annihilation of secondary photons is expected to
enhance the flux of $\bar\nu_e$ \cite{soeb}. These effects make the
discrimination between $p\gamma$ and $pp$ sources of neutrinos, by
detecting Glashow resonance events, very difficult. Exploring other
possibilities to reveal the underlying mechanisms of neutrino
production in astrophysical sources is still of great interest.
In our Galaxy the major sources of cosmic rays are SNRs and PWN.  High
energy gamma rays have been detected from many SNRs and PWN \cite{hess}
although the leading mechanism (hadronic or leptonic) of $\gamma$ ray
production in these sources is not yet known. They are also expected
to be sources of high energy neutrinos \cite{galneu}.  Protons and
other heavy nuclei are accelerated in diffusive shocks. They interact
with matter and radiation, lose energies and some of them escape from
the sources. $pp$ and $p\gamma$ interactions are the underlying
mechanisms of neutrino production in these sources as discussed
earlier. The gamma ray flux detected by CANGAROO \cite{can} from SN
1006 has been fitted with the theoretical 
model of $\pi^{0}$ decay \cite{ks}. The flux produced in inverse
Compton scattering of electrons by low energy photons is much lower
than that produced in $\pi^{0}$ decay. Recently, the evidence of
hadronic origin of gamma rays from Vela X \cite{vela} and PSR B1259-63
\cite{andrii} has been reported in two other papers. If a gamma ray
energy flux of $2\times10^{-13}~{\rm erg~cm^{-2}s^{-1}}$ is deteced from 
SN 1987A in the energy range of 0.1 to 10TeV then it would imply that
these gamma rays are of hadronic origin produced in $\pi^{0}$ decay
\cite{ber}.  The neutrino fluxes expected from various PWN were
calculated \cite{bed} considering detailed modeling of the sources and
found to be detectable in some cases by large scale neutrino
telescopes of $km^2$ area. High energy neutrino emission
from $p\gamma$ process inside the magnetospheres of isolated neutron
stars have been discussed for magnetars \cite{zhang}
and for young pulsars \cite{link}. In this paper the subject of our discussion is whether it is possible to constrain the parameter $x$, which denotes the fraction of the total number of Galactic protons participating in $p\gamma$ interactions. 
\section{Diffuse electron and positron spectra:}
Since high energy electrons and positrons lose energies faster than
other charged particles, by various energy loss mechanisms {\it e.g.}
bremsstrahlung, synchrotron emission and inverse Compton scattering,
their observed fluxes are most likely of Galactic origin. Above
1 GeV the dominant energy loss processes are inverse Compton
and synchrotron. It has been discussed in \cite{vb} that approximately
$10\%$ of electrons and all positrons are secondary particles produced
in nuclear interactions. There are some discrepancies in the measured 
data on electrons and positrons. Above several GeV, the large spread 
in the data can not be explained as an effect of solar modulation. 
Systematic errors in the measurements might be responsible for the 
disagreement in the data
taken by different experimental groups.  M\"uller \cite{mu} and Du
Vernois et al. \cite{du} suggested rescaling of the upper set of data
points to match the electron flux at 20 GeV obtained with the best fit
of the lower set of data points that include the recent results of
the AMS experiment \cite{ams}. After rescaling, all the data points in the
energy range of 3 GeV to 2 TeV can be fitted with a single power law
with the spectral index $3.44\pm0.03$ \cite{diego}.  With a similar
procedure of renormalization of the data from MASS 89 \cite{mass1} and
MASS 91 \cite{mass2} experiments, the positron data points can be
fitted with a single power law with the spectral index $3.43\pm 0.05$ 
above 0.7 GeV. The single power law fitted spectra are shown in Fig.4. 
and Fig.5. of \cite{diego}, and we have used their results in the present
paper.

\section{$p\gamma$, $pp$ interactions and $e^{\pm}$ fluxes:}

Cosmic rays with energy up to $10^{18}$ eV are most likely of
Galactic origin.  Protons can be accelerated to very high energies
inside the Galactic cosmic ray sources, {\it e.g.} SNRs, PWN, and
microquasars. One interesting feature is that interactions of 
shock accelerated protons with photons ($p\gamma\rightarrow 
e^{+}\nu_{\mu}\bar\nu_{\mu}\nu_en$) leading to the production 
of neutrinos generate high energy positrons, but not electrons. Most
other processes, on the other hand, produce equal number of 
electrons and positrons through pairs. These processes include
$p\gamma\rightarrow pe^{+}e^{-}$, $pp\rightarrow
e^{\pm}\nu_{\mu}\bar\nu_{\mu}\nu_e(\bar\nu_e)$ and
$\gamma\gamma\rightarrow e^{+}e^{-}$. Such an asymmetry makes
it possible to estimate the contribution of $p\gamma$ process
through evaluating the measured $e^{+}$ and $e^{-}$ fluxes.

The shock accelerated electron and proton spectra inside the sources
could be approximated as power laws in energy with  similar
spectral index, {\it i.e.} $S_{e^{-}}(E_e)=A_eE_{e}^{-\alpha}$ and
$S_p(E_p)=A_pE_{p}^{-\alpha}$. The possible values of 
$\alpha$ can be calculated by numerical simulations \cite{lem}.
 Electrons and positrons lose energy faster than protons inside the sources.
We have discussed in section 5 of this paper how the internal magnetic field and size of a source determine the escape probability of electrons and positrons from that source.
We denote the electron and positron spectra injected into the Galaxy as $Q_{e^{-}}(E_e)$ and $Q_{e^{+}}(E_e)$ respectively.
During propagation the cosmic ray protons and
nucleons would mainly interact with Galactic matter and produce 
secondary electrons and positrons, since the Galactic diffuse 
gamma-ray background is not high enough for significant positron
production by the propagating cosmic rays through $p\gamma$ 
interactions at the $\Delta$ resonance. The solution of
the transport equation for the propagation of $e^{\pm}$ in the Galaxy
is $N(E_e)\approx Q(E_e)V_{source}\tau(E_e)/V_{occ}(E_e)$, where
$Q(E_e)$ is the source flux, $V_{source}$ is the volume of the
sources, $V_{occ}(E_e)$ is the volume in the Galaxy occupied by
electrons or positrons of energy $E_e$, and $\tau(E_e)$ is the 
lifetime of the particle with energy $E_e$ \cite{vb}. 
We assume that out of the total number of protons inside all 
Galactic astrophysical objects, a fraction $x$ participates in 
$p\gamma$ interactions to produce neutrinos, another fraction 
$y$ participates in $pp$ interactions and the rest ($z$) escapes 
without any interactions or produce $e^{\pm}$ pairs in $p\gamma$ 
interactions, hence, $x+y+z=1$.  
The injection source fluxes of electrons and positrons can be 
expressed as
\beq
Q_{e^{-}}(E_{e})=P_{e^{-}}(E_{e})+F(E_{e})~,
\label{elec}
\eeq
\beq
Q_{e^{+}}(E_{e})=xf\phi_{p\gamma\rightarrow \nu+e^{+}}(E_{e})+F(E_{e})~.
\label{pos}
\eeq   
Here $F(E_{e})$ denotes the secondary electron or positron flux
injected into the Galaxy from all the interactions in which electron
and positron pairs are produced, including inside the astrophysical
sources and also during the propagation of the protons and nucleons in
the Galaxy. Rotation-powered pulsars inject electron-positron pairs
from the magnetosphere through magnetic one-photon ($\gamma B\rightarrow
e^{+}e^{-}$) pair production
processes \cite{rs75,dh82,zh00}. WIMP annihilations also generate 
electron-positron pairs \cite{tylka}. All these processes are included
in our $F(E_e)$ terms. We assume that
astrophysical sources contain protons and electrons, but positrons are
produced only in interactions.  Thus we denote $P_{e^{-}}(E_{e})$ as 
the primary diffuse electron flux directly injected from all the 
Galactic astrophysical objects (not generated through interactions). In
eqn(\ref{pos}), $\phi_{p\gamma\rightarrow \nu+e^{+}}(E_{e})$ is the
positron flux produced in photo-pion interactions normalized to one
pion per nucleon ($p\gamma\rightarrow\pi^{+}n\rightarrow
e^{+}\nu_{\mu}\nu_e\bar\nu_{\mu}n$) inside all astrophysical sources,
assuming that all protons participate in $p\gamma$ interactions to 
produce neutrinos. The positrons produced in this way lose energies by
bremsstrahlung, synchrotron and inverse Compton mechanisms before
their escape from the sources. The emitted fluxes are related to
the fluxes produced inside the sources by a factor $f \leq 1$.
This factor accounts for the escaping probability and energy losses of
electrons and positrons. The factor $x$ denotes the fraction of protons which
participate in $p\gamma$ interactions to produce neutrinos.  If $x=0$
there is no neutrino production through $p\gamma$ interactions inside
the sources, and $x=1$ if all the neutrinos are produced in $p\gamma$
interactions. One can than relate the observed and source fluxes of
electrons and positrons through the following ratios,
\bea
\frac{N_{e^{-}}(E_{e})-N_{e^{+}}(E_{e})}{N_{e^{-}}(E_{e})}
=\frac{Q_{e^{-}}(E_{e})-Q_{e^{+}}(E_{e})}{Q_{e^{-}}(E_{e})}
=\frac{P_{e^{-}}(E_{e})-xf\phi_{p\gamma\rightarrow \nu+e^{+}}(E_{e})}
{P_{e^{-}}(E_{e})+F(E_e)}
\label{ratio1}
\eea
 
Besides $p\gamma$ interactions, all other interactions
discussed earlier e.g. $pp$, $\gamma\gamma$, $\gamma B$, WIMP annihilations,
generate equal fluxes of electrons and positrons. The observed positron flux is about $10\%$ of the observed electron flux and they are expected to be produced in various interactions ($pp$, $p\gamma$ etc.) as discussed earlier. Hence, it is reasonable to assume that the flux of secondary electrons produced in
interactions inside astrophysical sources and inside the Galaxy during
the propagation of cosmic rays is much less compared to their original
flux which is not produced in any interaction \cite{vb}. From this
argument it follows that $F(E_e)<<P_{e^{-}}(E_e)$. 
 In $p\gamma$ interactions charged and neutral pions are produced with almost equal probabilities. Unlike positrons, the nucleons produced in $p\gamma$ 
interaction escape from the sources before losing their energies 
significantly. The charged pion carries about $20\%$ of the initial
proton's energy and the final state leptons share the
energy of $\pi^{+}$ equally. The energy $E_p$ of the proton and the
energy $E_e$ of the positron produced in one $p\gamma\rightarrow
\pi^{+}n\rightarrow e^{+}\nu_e\nu_{\mu}\bar\nu_{\mu}n$ interaction are
related as $E_e=\frac{1}{5}\times\frac{1}{4}E_p$. The total energies
carried by all the positrons of energy $E_e$ produced in $p\gamma$ interactions and all the protons of energy $E_p$ are related as \cite{wax},
\beq
E_e^2\phi_{p\gamma\rightarrow \nu+e^{+}}(E_{e})=\frac{1}{2}\times
\frac{1}{5}\times \frac{1}{4}{E_p^2}A_p{E_p^{-\alpha}}
\label{pgamma}
\eeq
where $A_pE_p^{-\alpha}$ is the shock accelerated proton number 
flux per unit energy (at $d E_p$) inside the sources of cosmic rays, and the
factor $1/2$ takes into account the contribution of protons that 
produce $\pi^0$ (and hence, not contributing to $e^{+}$ production).
The positron flux produced in $p\gamma$ interactions can be then
expressed as
\beq
\phi_{p\gamma\rightarrow \nu+e^{+}}(E_e)=\frac{A_p}{40}
\frac{E_p^{2-\alpha}}{E_e^2}
\eeq
 Eqn(\ref{ratio1}) 
can then be expressed in terms of spectral index $\alpha$, lepton energy $E_{e}$
,

\beq
\frac{N_{e^{-}}(E_{e})-N_{e^{+}}(E_{e})}{N_{e^{-}}(E_{e})}
=\frac{P_{e^{-}}(E_e)-xf\phi_{p\gamma\rightarrow \nu+e^+}(E_e)}
{P_{e^{-}}(E_e)+F(E_e)}\approx
\frac{P_{e^{-}}(E_e)}{P_{e^{-}}(E_e)+F(E_e)}-\frac{xf\phi_{p\gamma\rightarrow \nu+e^+}(E_e)}
{\Big[1+r(E_e)\Big]f A_eE_e^{-\alpha}}
\label{rt}
\eeq
where we have used $P_{e^{-}}(E_e)+F(E_e)=Q_{{e}^{-}}(E_e)$ from Eqn.(\ref{elec}) and the ratio of observed positron and electron fluxes \cite{diego,ams-01} is
  $r(E_e)=N_{e^{+}}(E_e)/N_{e^{-}}(E_e)$. $Q_{e^{-}}(E_e)$ can be greater than $fS_{e^{-}}(E_e)$ (the diffues electron flux emitted from all Galactic sources) at the most by a factor of $[1+r(E_e)]$. We have replaced $Q_{e^{-}}(E_e)$ with $[1+r(E_e)]fS_{e^{-}}(E_e)$ in the denominator of the second term on the right hand side of Eqn.(\ref{rt}).  The positron's energy $E_e$ and the proton's energy $E_p$ are related as $E_{e}=E_p/20$. Assuming that the shock accelerated electrons of mass $m_e$ and protons of mass $m_p$ have the same values of spectral indices the
ratio of their shock accelerated fluxes at relativistic 
energies $(E>m_pc^2)$ is
$\frac{A_e}{A_p}=\zeta\Big(\frac{m_e}{m_p}\Big)^{(\alpha-1)/2}$
\cite{vb}, 
where $\zeta$ is the ratio of the non relativistic electron and
proton fluxes at any energy $E_k<m_ec^2$ (for non relativistic particles it is the kinetic energy). The numbers of non relativistic primary electrons and protons inside the sources are expected to be equal hence, $\zeta \sim 1$. 
 We can then constrain $x$ using the following 
expression,
\beq
{x}\approx40\Big[1+r(E_e)\Big]\times{\Big(\frac{1}{20}\Big)}^{2-\alpha} 
\Big(\frac{m_e}{m_p}\Big)^{\frac{\alpha-1}{2}}
\Big[r(E_{e})-\frac{F(E_e)}{P_{e^{-}}(E_e)+F(E_e)}\Big]~.
\label{x}
\eeq
The ratio of observed positron and electron fluxes $r(E_e)$ is available from observed data. However, we do not know how much is the ratio $F(E_e)/[P_{e^{-}}(E_e)+F(E_e)]$. Assuming $F(E_e)=0$, one can put an upper bound on $x$.

\beq
{x}\leq 40\Big[1+r(E_e)\Big]\times{\Big(\frac{1}{20}\Big)}^{2-\alpha}
\Big(\frac{m_e}{m_p}\Big)^{\frac{\alpha-1}{2}}r(E_e)~.
\label{ub}
\eeq
Thus it is possible to constrain $x$ using the observed data on positron and electron fluxes.
\section{Results:}
The current data on cosmic ray positron flux is up to about 60 GeV.
 In \cite{ams-01} the data are given upto 30 GeV. Below 10 GeV solar modulation effect becomes important hence we have considered the energy range above this energy. As the ratio of the measured fluxes of positrons and electrons is small the errors involved in current measurements of this ratio affect the result. In future with more precise measurements the flux ratio could be determined more accurately. The average value of the observed flux ratios is 0.1.
We calculate the upper bound on $x$ for an energy range of 10GeV to 60GeV using
 the power law fit to the compilation of data from \cite{diego}. Our result is
$x\leq0.1$ for $\alpha=2.1-2.4$. The upper limit on $x$ slightly decreases with increasing $\alpha$. In general the factor $x$ depends on energy, but in the 10-60 GeV range it is insensitive to the energy if we use the power law fit to the compiled data from \cite{diego}. In future it would be possible to constrain $x$ at higher energies when the positron fluxes at higher energies are available.
   
\section{Discussions and Conclusion:}
In order to contribute to the observed positron flux at earth, positrons
must not cool before escaping the source. This requires that the cooling
time scale $t_{cool}$ is longer than the escaping time scale $t_{esc}$.
For standard synchrotron cooling, the cooling time scale can be
estimated as 
\beq
t_{cool}=\frac{m_e\gamma_ec^2}{\frac{4}{3}\sigma_T 
c\gamma_e^2\frac{B^2}{8\pi}},
\label{cooling}
\eeq
where $m_e$ is mass of a positron, $c$ is the speed of light, 
$\gamma_e$ is Lorentz factor of the positron, $B$ is the internal 
magnetic field of the source, and $\sigma_T$ is Thomson cross section. 
To estimate the escaping time scale, we adopt the random walk
approximation. In a random magnetic field, the mean free path
for each scattering may be estimated as the Larmor radius
$R_L$. The number of scattering of a positron before escaping the 
source is $\sim (\frac{R}{R_L})^2$, where $R$ is the radius of the 
particle accelerating region inside a source. We therefore have
\beq
t_{esc}=\Big(\frac{R}{R_L}\Big)^2\frac{R_L}{c}~.
\label{escape}
\eeq
Since $R_L=\frac{m_e\gamma_ec^2}{eB}$, it is interesting to observe that
$t_{cool}$ and $t_{esc}$ have the same $\gamma_e$ dependence.
This suggests that for a particular source, either all positrons
escape the source ($t_{cool}>t_{esc}$) or all positrons cool before
escaping ($t_{cool}<t_{esc}$). With eqn.(\ref{escape}) and 
eqn.(\ref{cooling}), the positron-escaping condition is
\beq
B_{-3}\le3.4 R_{15}^{-2/3}
\label{magsize}
\eeq 
where the convention $Q_i=Q/10^i$ is adopted in cgs units.
The typical values of $B$ required to explain the observed radio 
and x-ray data from SN 1006 and SN 1987A are about $150{\mu}$G 
\cite{ks} and $10$mG \cite{ber}, respectively. The size of the
positron emission region may vary from sources to sources.
We can see that cosmic sources parameters are marginal to
satisfy the constraint. 
Generally we expect some sources can satisfy eqn. (\ref{magsize})
and contribute to the observed diffuse positron spectra. 
The factor $f$ therefore denotes the fraction of all the $p\gamma$ 
positron generation sources that satisfy eqn.(\ref{magsize}). 
The upper bound on the fraction of Galactic protons participating in $p\gamma$ interactions in our Galaxy is found to be $10\%$. Our current result is relevant to the positron energy range between $10$ GeV to $60$ GeV. In  
future it would be possible to extend our results 
with more observational data on positron fluxes at higher energies.   
\section{Acknowledgement}    
This work was supported by NASA under grants NNG05GB67G and NNG06GH62G.


\begin{thebibliography}{99}
\bibitem{amanda1}AMANDA Coll., M. Ackermann et al., Phys.Rev. D {\bf 71}, 077102 (2005).
\bibitem{baikal}Baikal Coll., R. Wischnewski et al., Int.J.Mod.Phys. A {\bf 20}, 6932(2005).
\bibitem{antares}ANTARES Coll., J. A. Aguilar, et al., astro-ph/0606229.
\bibitem{K3net}KM3NeT Coll., U. F. Katz, astro-ph/0606068. 
\bibitem{ice1}http://icecube.wisc.edu/.
\bibitem{rice}RICE Coll., I. Kravchenko et al., Phys. Rev. D {\bf 73}, 082002 (2006).
\bibitem{anita}ANITA Coll., S. W. Barwick et al., Phys.Rev.Lett. {\bf 96}, 171101(2006).
\bibitem{lofar}H. J. A. R\"ottgering, New Astron.Rev. {\bf 47}, 405 (2003).
\bibitem{amanda2}AMANDA Coll., J. Ahrens et al., Phys.Rev.Lett. {\bf 92}, 071102 (2004).
\bibitem{amanda3}AMANDA Coll., A. Achterberg et al., {Proc. of $29^{th}$ International Cosmic Ray Conference}, Aug (2005), Pune, India, astro-ph/0509330.
\bibitem{ice2}IceCube Coll., Mathieu Ribordy, et al., {Invited talk in $5^{th}$ International Conference on Non-accelerator New Physics  (NANP05)}, Jun. (2005) Dubna, Russia, astro-ph/0509322.
\bibitem{strong}I. V. Moskalenko, A. W. Strong, ApJ {\bf 493}, 694 (1998).
\bibitem{anch}L. A. Anchordoqui, H. Goldberg, F. Halzen, T. J. Weiler, Phys. Lett. B {\bf 621}, 18 (2005).
\bibitem{glash}S. L. Glashow, Phys. Rev. {\bf 118}, 316 (1960).
\bibitem{pijush}P. Bhattacharjee, N. Gupta, hep-ph/0501191.  
\bibitem{kashti}T. Kashti, E. Waxman, Phys. Rev. Lett. {\bf 95}, 181101 (2005).
\bibitem{soeb}S. Razzaque, P. M\'esz\'aros, E. Waxman, Phys. Rev. D {\bf 7}3, 103005 (2006).
\bibitem{hess}The Hess Collab., F. Aharonian et al., ApJ {\bf 636}, 777(2006);
A\&A {\bf 449}, 223 (2006).
\bibitem{galneu}M. L. Costantini, F. Vissani, Astropart. Phys. {\bf 23}, 447 (2005); M. D. Kistler, J. F. Beacom, Phys. Rev. D {\bf 74}, 063007 (2006); F. Vissani, Astropart. Phys. {\bf 26}, 310 (2006); A. Kappes, J. Hinton, C. Stegmann, F. Aharonian, astro-ph/0607286.
\bibitem{can}http://icrhp9.icrr.u-tokyo.ac.jp/.
\bibitem{ks}L.T.Ksenofontov, E.G.Berezhko, H.J.V\"olk, A\&A, {\bf 443}, 973 (2005).
\bibitem{vela}D. Horns, F. Aharonian, A. Santangelo, A.I.D. Hoffmann, C. Masterson , A \& A, {\bf 451}L, 51H (2006). 
\bibitem{andrii}A. Neronov, M. Chernyakova, {Proc. of The Multi Messenger Approach to High-Energy Gamma-Ray Sources}, Barcelona, June 2006, astro-ph/0610139.
\bibitem{ber}E.G. Berezhko, L.T. Ksenofontov, ApJ Lett. in press (2006), astro-ph/0608586.
\bibitem{bed}W. Bednarek, R. J. Protheroe, Phys. Rev. Lett. {\bf 79}, 2616 (1997); W. Bednarek, A\&A {\bf 407}, 1 (2006).
\bibitem{zhang}B. Zhang, Z. G. Dai, P. Meszaros, E. Waxman, A. K. Harding, ApJ {\bf 595}, 346 (2003). 
\bibitem{link}B. Link, F. Burgio, MNRAS {\bf 371}, 375 (2006).
\bibitem{vb}V. S. Berezinsky et al., {\it Astrophysics of Cosmic Rays, published by North-Holland (1990)}.
\bibitem{mu}D. M\"uller, Adv. Space Res. {\bf 27}, 659 (2001).
\bibitem{du}M. A. Du Vernois et al., ApJ {\bf 559}, 296 (2001).
\bibitem{ams}J. Alcaraz, B. Alpat, G. Ambrosi, Phys. Lett. B {\bf 484}, 10 (2000).
\bibitem{diego}D. Casadei, V. Bindi, ApJ {\bf 612}, 262 (2004).
\bibitem{mass1}R. L. Golden et al., ApJ {\bf 436}, 769 (1994).
\bibitem{mass2}C. Grimani et al., A\& A {\bf 392}, 287 (2002).
\bibitem{lem}M. Lemoine, G. Pelletier, ApJ {589}, L73 (2003).
\bibitem{rs75}M. A. Ruderman, P. G. Sutherland, ApJ {\bf 196}, 51 (1975).
\bibitem{dh82}J. K. Daugherty, A. K. Harding, ApJ {\bf 252}, 337 (1982).
\bibitem{zh00}B. Zhang, A. K. Harding, ApJ {\bf 532}, 1150 (2000).
\bibitem{tylka}A. J. Tylka, Phys. Rev. Lett. {\bf 63}, 840 (1989).
\bibitem{wax}E. Waxman, J. N. Bahcall, Phys.Rev. D {\bf 59}, 023002 (1999).
\bibitem{ams-01}M. Aguilar et al., Phys. Lett. B {\bf 646}, 145 (2007).
\end{thebibliography}
\end{document}